\documentclass[showpacs,aps,prb,twocolumn]{revtex4}

\usepackage{graphicx}

\def  \bsig    {\mbox{\boldmath$\sigma$}}

\begin{document}

\title{Graphene in periodic deformation fields: dielectric screening and plasmons}
\author{V.~K.~Dugaev}
\affiliation{Institut f\"ur Physik,
Martin-Luther-Universit\"at Halle-Wittenberg,
Heinrich-Damerow-Str. 4, 06120 Halle, Germany, and\\
Department of Physics,
Rzesz\'ow University of Technology, Al.~Powsta\'nc\'ow Warszawy 6,
35-959 Rzesz\'ow, Poland, and\\
Department of Physics and CFIF, Instituto Superior T\'ecnico,
Universidade T\'ecnica de Lisboa, Av. Rovisco Pais, 1049-001 Lisbon, Portugal}
\author{M. I. Katsnelson}
\affiliation{Radboud University Nijmegen, Institute for Molecules and Materials,
Heyendaalseweg 135, 6525 AJ Nijmegen, The Netherlands}

\begin{abstract}
We consider the effect of periodic scalar and vector potentials
generated by periodic deformations of the graphene crystal lattice,
on the energy spectrum of electrons. The dependence of electron velocity
near the Dirac point on the periodic perturbations
of different types is discussed.
We also investigated the effect of screening of the scalar potential by
calculating the dielectric function as a function of the wave length
of the periodic potential. This calculation shows that the periodic
scalar field is strongly suppressed by the screening.
Using the dependence of electron velocity on the periodic
field we also studied the variation of the plasmon spectra in graphene.
We found that the spectrum of plasmon excitations
can be effectively controlled by the periodic strain field.
\end{abstract}

\date{\today }

\maketitle


\section{Introduction}

The enormous interest to graphene is related to the unique
physical properties of this two-dimensional material
\cite{geim07,castro09,katsbook}, which most possibly will be used
in the future in numerous technological applications \cite{acp,geim09,bonnacorso10,novoselov11}.
An example of such properties
of interest for the applications is the existence of very unusual
spectrum of plasmon excitations with THz frequencies \cite{jablan09,dubinov11},
that can be used in optoelectronics and communications.

One of the most important problems to be solved to use graphene in
electronics is the realization of effective
control of the parameters of energy spectrum such as
the electron energy gap and/or the velocity of electron and holes.
It is already known that by using the electrostatic gating one can
vary the carrier density of graphene (i.e., the location of chemical potential).
Recently, it was also proposed to use the external strain field
to change the energy spectrum -- this is called the strain
engineering of graphene \cite{katsbook,pereira09,guineaNP10,guinea10}.
The idea is mostly based on unusually strong effect of the external deformation
acting on the energy spectrum of graphene quite similar to the
external electric and magnetic fields
\cite{katsbook,vozmediano10}.

In this work we consider the effect of periodic fields, which
can be generated by periodic deformations, on the energy spectrum,
screening of electron-electron interaction, and on the plasmon excitations
in graphene. It was already
pointed out that such periodic fields do not open the gap
near the Dirac point but affect the velocity parameter
of electrons and holes \cite{park08,park08_N,tan10}
making the energy spectrum anisotropic. Recently, the effect of periodic modulation
on electron spectrum of graphene has been studied experimentally \cite{LeRoy}.
Here we reconsider this problem in more details, concentrating on
possible coexistence of periodic scalar and vector potentials, which, to our
knowledge, has not been done before. This is important since a generic deformation
produces both scalar and vector potentials \cite{katsbook,guinea10,vozmediano10}.
We use a different method to solve the problem \cite{com1},
and find that our numerical results are in
agreement with those of Refs.~[\onlinecite{park08,tan10}].
We also discuss the role of screening due to the free carriers
in graphene, and we find that screening can substantially suppress the
effect of periodic scalar potential.

The combined effect of periodic scalar potential and constant magnetic
field in graphene has been recently investigated by Wu et al.\cite{wu12}
They found that the structure of Landau levels can be also strongly
affected by the one-dimensional (1D) periodic fields.

The physics of plasmons in graphene has been intensively discussed
recently by many authors.
\cite{hwang07,sarma09,tudorovskiy10,nikitin11,abedinpour11}
The calculations have been performed in frame of standard RPA
approximation taking into account the electron energy structure of
graphene near the Dirac points, as well as for the whole
energy spectrum at the honeycomb lattice \cite{yuan11}.
The effects of magnetic field, finite temperature, chemical potential
have been investigated in the same approach \cite{pyatkovskiy11}.
In this work we discuss the effect of external periodic fields on the
screening and on the energy spectrum of plasmons.

\section{Electron energy spectrum in periodic fields}

We consider first the transformation of electron energy spectrum
related to deformations of the graphene lattice.
It is known that the deformation of graphene is equivalent to
the generation of electric
and magnetic fields, which can be described by scalar $V({\bf r})$
and vector ${\bf A}({\bf r})$ potentials \cite{katsbook,vozmediano10}. The relations between
the components of strain tensor $u_{ij}({\bf r})$ and the scalar
and vector gauge potentials are \cite{suzuura02,manes07}
\begin{eqnarray}
\label{1}
&&V({\bf r})=g\, (u_{xx}+u_{yy}),
\nonumber \\
&&A_x({\bf r})=\frac{\beta t}{a_0}\, (u_{xx}-u_{yy}),
\hskip0.2cm
A_y({\bf r})=-\frac{2\beta t}{a_0}\, u_{xy},
\end{eqnarray}
where $g$ is the deformation potential, $t$ is the hopping energy,
the parameter $\beta $ is defined by $\beta =-\partial \log t/\partial \log a_0$,
and $a_0$ is the lattice constant.
In the following we consider one-dimensional periodicity of the deformation,
and in view of Eq.(1) we assume that the periodic in $x$ strain field
generates periodic scalar and vector potentials $V(x)$ and ${\bf A}(x)$.

The Hamiltonian of electrons in graphene near the $\mathcal{K}$ Dirac point in
periodic scalar and vector fields reads
\begin{eqnarray}
\label{2}
\mathcal{H}=-iv\bsig \cdot (\nabla -i{\bf A})+V,
\end{eqnarray}
where the Pauli matrices $\bsig $ act on the sublattice label and we use the units $\hbar = e =1$.
This Hamiltonian describes low-energy excitations of the electronic system
in graphene.

The corresponding Schr\"odinger equation for spinor wave function
$\psi ^T({\bf r})=(\varphi , \chi )$ is
\begin{eqnarray}
\label{3}
\left( \begin{array}{cc}
\varepsilon -V & iv\partial _- +vA_- \\
iv\partial _+ +vA_+ & \varepsilon - V
\end{array} \right)
\left( \begin{array}{c} \varphi \\ \chi \end{array} \right) =0,
\end{eqnarray}
where $\partial _\pm =\partial _x\pm i\partial _y$ and $A_\pm =A_x\pm iA_y$.
Since the potentials ${\bf A}(x)$ and $V(x)$ do not depend on $y$
and depend periodically on $x$,
we take $\varphi ,\chi \sim e^{i{\bf k}\cdot {\bf r}}$.

In frame of the $k\cdot p$ approximation \cite{tsidilkovskii}, we have to calculate first the
wavefunction $\psi (x)$ at ${\bf k}=0$.
The corresponding equations for the spinor components at ${\bf k}=0$ are
\begin{eqnarray}
\label{4}
(\varepsilon -V)\varphi +iv\chi '+vA_-\chi =0,
\\
iv\varphi '+vA_+\varphi +(\varepsilon -V)\chi =0.
\end{eqnarray}
From  Eq.~(5) follows
\begin{eqnarray}
\label{6}
\chi =-\frac{iv\varphi '}{\varepsilon -V}-\frac{vA_+\varphi }{\varepsilon -V}\, ,
\end{eqnarray}
where prime means $\partial _x$. Substituting Eq.(6) into Eq.(4) we obtain the following equation for
$\varphi (x)$
\begin{eqnarray}
\label{7}
(\varepsilon -V)\varphi
+\frac{v^2\varphi ''}{\varepsilon -V}
+\frac{v^2V'\varphi '}{(\varepsilon -V)^2}
-\frac{iv^2A'_+\varphi }{\varepsilon -V}
-\frac{iv^2A_+\varphi '}{\varepsilon -V}
\nonumber \\
-\frac{iv^2V'A_+\varphi }{(\varepsilon -V)^2}
-\frac{iv^2A_-\varphi '}{\varepsilon -V}-\frac{v^2A_+A_-\varphi }{\varepsilon -V}=0.
\end{eqnarray}

Let us assume the existence of solutions of Eq.~(7) with $\varepsilon =0$
(here we assume that $V\ne 0$).
Then we get
\begin{eqnarray}
\label{8}
\varphi ''-\left( \frac{V'}{V}+2iA_x\right) \varphi '
+\left( \frac{V^2}{v^2}-iA'_++\frac{iV'A_+}{V}
\right. \nonumber \\ \left.
-A_+A_-\right) \varphi =0.
\end{eqnarray}
This is the equation for $\varphi (x)$ in the $\mathcal{K}$ Dirac point,
corresponding to the lowest energy band.

Let us consider first some particular cases, when the periodic field
is purely scalar $V(x)$ or purely vector field ${\bf A}(x)$.

\subsection{Periodic scalar potential}

In the case when ${\bf A}(x)=0$ and $V(x)\ne 0$, Eq.(8) essentially
simplifies to
\begin{eqnarray}
\label{9}
\varphi ''-\frac{V'}{V}\, \varphi '
+\frac{V^2}{v^2}\, \varphi =0
\end{eqnarray}
and has two different solutions
\begin{eqnarray}
\label{10}
\varphi _{1,2}(x)=\exp \left( \pm \frac{i}{v}\int _0^xV(x')\,
dx'\right) .
\end{eqnarray}
Correspondingly, using Eqs.(6) and (10) we obtain
\begin{eqnarray}
\label{11}
\chi _{1,2}(x)=\mp \exp \left( \pm \frac{i}{v}\int _0^xV(x')dx'\right) .
\end{eqnarray}
Then the normalized basis functions in $k\cdot p$ approximation are
\begin{eqnarray}
\label{12}
\psi _1({\bf r})=\frac{e^{i{\bf k\cdot r}}}{\sqrt{2S}}
\left( \begin{array}{c} \varphi _1\\ \chi _1\end{array} \right) ,
\hskip0.3cm
\psi _2({\bf r})=\frac{e^{i{\bf k\cdot r}}}{\sqrt{2S}}
\left( \begin{array}{c} \varphi _2 \\ \chi _2\end{array} \right) ,
\end{eqnarray}
where $S={\cal L}_x{\cal L}_y$ is the area of graphene sample.
Calculating the matrix elements of the Hamiltonian (2) with
basis functions (12) we find the effective Hamiltonian
\begin{eqnarray}
\label{13}
\tilde{\mathcal{H}}=\left( \begin{array}{cc}
-vk_x & v(\gamma _1k_x-\gamma _2k_y) \\
v(\gamma _1k_x-\gamma _2k_y) & vk_x ,
\end{array}\right) ,
\end{eqnarray}
where we denote
\begin{eqnarray}
\label{14}
\gamma _1=\frac1{L}\int _0^{L}dx\,
\cos \left( \frac2{v}\int _0^{x}V(x')\, dx'\right) ,
\\
\gamma _2=\frac1{L}\int _0^{L}dx\,
\sin \left( \frac2{v}\int _0^{x}V(x')\, dx'\right) ,
\end{eqnarray}
and $L$ is the period of the potential $V(x)$.
In correspondance with Eqs.(14) and (15) both parameters $\gamma _1,\gamma _2<1$.

The Hamiltonian (13) describes low-energy spectrum in the
periodic scalar field.
It has the following eigenvalues
\begin{eqnarray}
\label{16}
\varepsilon _{1,2}({\bf k})=\pm v\sqrt{k_x^2+(\gamma _1k_x-\gamma _2k_y)^2} .
\end{eqnarray}
Taking on alternate $k_y=0$ and $k_x=0$ we find that
due to the periodic scalar field, the components of
electron velocity in directions $x$ and $y$, are renormalized,
respectively, as
$\tilde{v}_x/v=\sqrt{1+\gamma _1^2}$ and $\tilde{v}_y/v=|\gamma _2|$.
Thus, in this case we always obtain $v_x>v$ and $v_y<v$.

\subsection{Periodic vector potential}

In the case when $V(x)=0$ and ${\bf A}(x)\ne 0$ we can use Eqs.(4) and (5)
to find directly from these equations the spinor components of the wave function
in the $\mathcal{K}$ point, ${\bf k}=0$
\begin{eqnarray}
\label{17}
\varphi (x)=\exp \left( i\int _0^xA_+(x')\, dx'\right),
\nonumber \\
\chi (x)=\exp \left( i\int _0^xA_-(x')\, dx'\right) .
\end{eqnarray}
Then we can introduce two $k\cdot p$ basis functions in the form
\begin{eqnarray}
\label{18}
\psi _1({\bf r})=N_1e^{i{\bf k\cdot r}}
\left( \begin{array}{c} \varphi \\ 0\end{array} \right) ,
\hskip0.2cm
\psi _2({\bf r})=N_2e^{i{\bf k\cdot r}}
\left( \begin{array}{c} 0 \\ \chi \end{array} \right) ,
\end{eqnarray}
where $N_i$ are the normalization factors
\begin{eqnarray}
\label{19}
N_{1,2}
=\left[ \mathcal{L}_y
\int _0^{\mathcal{L}_x}\exp \left(\mp 2\int _0^{x}A_y(x')\, dx'\right) dx\right] ^{-1/2}.
\hskip0.3cm
\end{eqnarray}
for the crystal of size $\mathcal{L}_x\times \mathcal{L}_y$.
Calculating the matrix elements of the Hamiltonian (2) with $V(x)=0$ in the
basis of $k\cdot p$ functions (18) we get
\begin{eqnarray}
\label{20}
\tilde{\mathcal{H}}=\left( \begin{array}{cc}
0 & \tilde{v}k_- \\
\tilde{v}^*k_+ & 0
\end{array} \right) ,
\end{eqnarray}
where
\begin{eqnarray}
\label{21}
\tilde{v}=vN_1N_2\mathcal{L}_y
\int _0^{\mathcal{L}_x} \varphi ^*(x)\, \chi (x)\, dx.
\end{eqnarray}

\begin{figure}[ptb]
\vspace*{-0.6cm}
\hspace*{-1cm}
\includegraphics[width=1.1\linewidth]{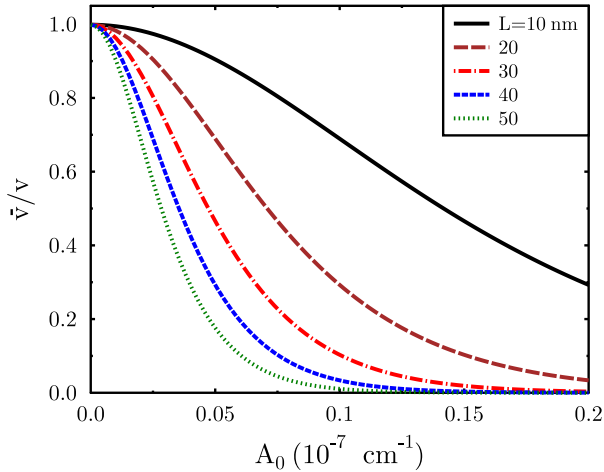}
\caption{Dependence of the renormalized velocity factor $\tilde{v}/v$ on
the amplitude $A_0$ of periodic potential $A_y(x)$ at different values of
periodicity parameter $L$.}
\end{figure}

Using Eqs.(17) and (21) we finally obtain
\begin{eqnarray}
\label{22}
\frac{\tilde{v}}{v}
=\Big[ \frac1{L^2}\int _0^{L}
\exp \left( -2\int _0^{x_1}A_y(x')\, dx'\right) dx_1\hskip0.5cm
\nonumber \\ \times
\int _0^{L}
\exp \left( 2\int _0^{x_2}A_y(x')\, dx'\right) dx_2
\Big] ^{-1/2}.
\end{eqnarray}
The dependence of $\tilde{v}/v$ from the amplitude $A_0$ of the
periodic potential $A_y(x)=A_0\sin (2\pi x/L)$ at different
values of period $L$ is presented in Fig.~1. The renormalized electron
velocity decreases in the periodic vector field.
This is in agreement\cite{com1} with Ref.~[\onlinecite{tan10}].

\subsection{General case: both scalar and vector potentials are nonzero}

In the general case when both $V(x)\ne 0$ and ${\bf A}(x)\ne 0$ we cannot
find simple analytic solutions but we can analyze further Eq.(8)
for $\varphi (x)$. For this purpose we present this equation as
\begin{eqnarray}
\label{23}
\varphi ''+a\varphi '+b\varphi =0,
\end{eqnarray}
where we denoted
\begin{eqnarray}
\label{24}
&&a(x)=-\frac{V'}{V}-2iA_x,
\\
&&b(x)=\frac{V^2}{v^2}-iA'_++\frac{iV'A_+}{V}-A_+A_-.
\end{eqnarray}
Then after substitution
\begin{eqnarray}
\label{26}
\varphi (x)=f(x)\, \exp \left( -\frac12 \int _0^x a(x')\, dx'\right)
\end{eqnarray}
we obtain the equation for the function $f(x)$
\begin{eqnarray}
\label{27}
-\frac12\, f''+\Big( \frac{a'}4+\frac{a^2}8 -\frac{b}2 \Big) f=0.
\end{eqnarray}
This is the Schr\"odinger equation for a particle of unit mass with energy
$\varepsilon =0$ in the potential
\begin{eqnarray}
\label{28}
U(x)
=-\frac{V''}{4V}+\frac{3(V')^2}{8V^2}+\frac{A_y^2-A'_y}2+\frac{V'A_y}{2V}
-\frac{V^2}{2v^2}.\hskip0.3cm
\end{eqnarray}
Here we note that $U(x)$ does not depend on $A_x$. It can be used
when $A_y=0$ since in this case we can take the solution
for $f(x)$ corresponding to $\varphi (x)$ from Eq.~(10). This way
we obtain simple generalization of Eqs.~(10) and (11) for
$A_x\ne 0$ and $A_y=0$
\begin{eqnarray}
\label{29}
&&\varphi _{1,2}(x)
=\exp \left[ i\int _0^x\left( \pm \frac{V(x')}{v}+A_x(x')\right) \, dx'\right] ,
\\
&&\chi _{1,2}(x)
=\mp \exp \left[ i\int _0^x\left( \pm \frac{V(x')}{v}+A_x(x')\right) \, dx'\right] .
\hskip0.7cm
\end{eqnarray}

Turning back to Schr\"odinger equation (27)
we also note that the potential (28)
is real and periodic, $U(x)=U(x+L)$, so that we can present it as
\begin{eqnarray}
\label{31}
U(x)=\sum _{n=1}^N\left( u_n e^{2\pi inx/L}+u_n^* e^{-2\pi inx/L}\right) ,
\end{eqnarray}
where the coefficients $u_n$ can be found from the specific shape of
potential $U(x)$.

We are looking for a periodic solution $f(x)$ of Eq.(27), which can be
presented as
\begin{eqnarray}
\label{32}
f(x)=\sum _m f_m e^{2\pi imx/L}
\end{eqnarray}
with $m$ is integer.
Then using Eqs.(31), (32) and (27) we find the matrix equation for the
coefficients $f_m$ in Eq.(32)
\begin{eqnarray}
\label{33}
\sum _mA_{nm}f_m=0,
\end{eqnarray}
where $A_{nm}=\frac12 \, k_n^2 \delta _{nm}+u_{n-m}+u^*_{m-n}$ with
$k_n=2\pi n/L$ and $u_n=0$ for any $n<1$.

One can also look for the solution of Eq.(8) in the form $\varphi (x)=e^{is(x)}$.
This representation can be more convenient for numerical calculations
with arbitrary periodic functions $V(x)$ and ${\bf A}(x)$.
In this approach we get the first order differential equation for
$\xi (x)$
\begin{eqnarray}
\label{34}
\xi '+i\xi ^2-\left( \frac{V'}{V}+2iA_x\right) \xi
-\frac{iV^2}{v^2}-A'_++\frac{V'A_+}{V}
\nonumber \\
+iA_+A_-=0,
\end{eqnarray}
where $\xi (x)=s'(x)$.
Note that the transition to the case $V(x)\to 0$ formally corresponds
to $V'/V\to \infty $ in Eq.(34).

One can assume that like Eqs.(12) and (18),
in the general case there are also two solutions of Eq.(8)
for the envelope function $\psi (x)$.
Then there are also two different solutions
of Eq.(34), $\xi _1(x)$ and $\xi _2(x)$.
Correspondingly we get two solutions for the first spinor component
\begin{eqnarray}
\varphi _i(x)=\exp \left( i\int _0^x \xi _i(x')\, dx' \right) ,
\hskip0.5cm i=1,2.
\nonumber
\end{eqnarray}
Then using Eq.(6) with $\varepsilon =0$
we find the other components $\chi _1(x)$ and $\chi _2(x)$.
The obtained spinor function
$(\varphi _i,\, \chi _i)^T$ should be properly normalized.
As before, we use these independent solutions for our $k\cdot p$ basis presented
by $\psi _i({\bf r})=N_ie^{i{\bf k}\cdot {\bf r}}(\varphi _i, \chi _i)^T$.

Thus, in the general case of arbitrary periodic perturbations
we obtain the effective Hamiltonian
\begin{eqnarray}
\label{35}
\tilde{\mathcal{H}}=\left( \begin{array}{cc}
2v\alpha _1k_x+2v\alpha_2 k_y & v\gamma k_-+v\delta k_+ \\
v\gamma ^*k_++v\delta ^*k_- & 2v\beta _1k_x+2v\beta _2k_y
\end{array} \right) ,
\end{eqnarray}
where we denote
\begin{eqnarray}
\label{36}
&&\alpha _1+i\alpha _2
=N_1^2\mathcal{L}_y\int _0^{\mathcal{L}_x}\varphi _1^*\, \chi _1\, dx,
\nonumber \\
&&\beta _1+i\beta _2
=N_2^2\mathcal{L}_y\int _0^{\mathcal{L}_x}\varphi _2^*\, \chi _2\, dx,
\nonumber \\
&&\zeta
=N_1N_2\mathcal{L}_y\int _0^{\mathcal{L}_x}\varphi _1^*\, \chi _2\, dx,
\\
&&\delta
=N_1N_2\mathcal{L}_y\int _0^{\mathcal{L}_x}\chi _1^*\, \varphi _2\, dx.
\nonumber
\end{eqnarray}
The eigenvalues of Hamiltonian (35) are
\begin{eqnarray}
\label{37}
\varepsilon _{1,2}({\bf k})
=v(\alpha _1+\beta _1)k_x+v(\alpha _2+\beta _2)k_y\hskip2cm
\nonumber \\
\pm v\big[(\alpha _1-\beta _1)^2k_x^2+(\alpha _2-\beta _2)^2k_y^2
+(|\zeta |^2+|\delta |^2)k^2
\nonumber \\
+2\, {\rm Re}\, (\zeta \delta ^*)(k_x^2-k_y^2)
+4\, {\rm Im}\, (\zeta \delta ^*)k_xk_y\big] ^{1/2}.\hskip0.3cm
\end{eqnarray}
For $k_x=0$ we obtain
$\varepsilon _{1,2}(k_y)=\tilde{v}^y_{1,2}k_y$,
where
\begin{eqnarray}
\label{38}
\tilde{v}^y_{1,2}/v=(\alpha _2+\beta _2)
\pm \big[ (\alpha _2-\beta _2)^2
+(|\zeta |^2+|\delta |^2)
\nonumber \\
-2\, {\rm Re}\, (\zeta \delta ^*)\big] ^{1/2},
\end{eqnarray}
and for $k_y=0$ we get
$\varepsilon _{1,2}(k_x)=\tilde{v}^x_{1,2}k_x$,
where
\begin{eqnarray}
\label{39}
\tilde{v}^x_{1,2}/v=(\alpha _1+\beta _1)
\pm \big[ (\alpha _1-\beta _1)^2
+(|\zeta |^2+|\delta |^2)
\nonumber \\
+2\, {\rm Re}\, (\zeta \delta ^*)\big] ^{1/2}.
\end{eqnarray}
Using Eqs.(36), (38), and (39) one finds the components of
electron velocity in graphene in the case of arbitrary periodic
perturbation described by the fields $V(x)$ and ${\bf A}(x)$.

\subsection{Longitudinal standing wave}

For a longitudinal strain wave, the components of deformation are $u_x=u_x(x)$
and $u_y=0$. Then in accordance with Eq.(1) we get $A_y=0$ and due to
Eq.(28) we can use the solutions (29) and (30).

\begin{figure}[ptb]
\vspace*{-0.6cm}
\hspace*{-1cm}
\includegraphics[width=1.1\linewidth]{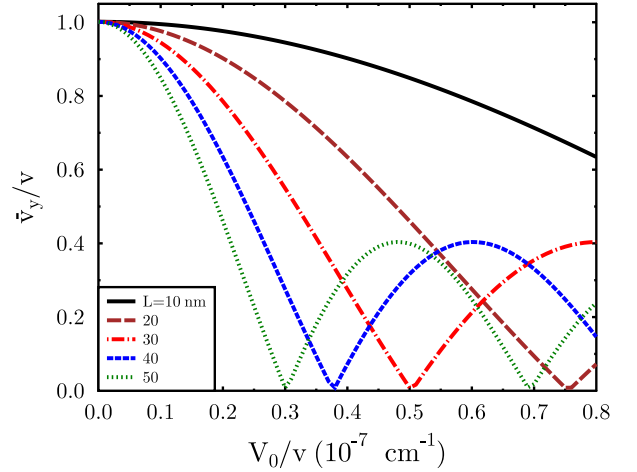}
\caption{Dependence of the renormalized velocity factor $\tilde{v_y}/v$ on
the amplitude of periodic scalar potential $V(x)$ at different values of
periodicity parameter $L$.}
\end{figure}

Using Eq.(36) we find
\begin{eqnarray}
\label{40}
&&\alpha =-1/2,\hskip0.5cm \beta =1/2,
\nonumber \\
&&\zeta =-\delta
=\frac1{2L}\int _0^{L}
\exp \left( -\frac{i}{v}\int _0^x V(x')\, dx'\right) dx,
\hskip0.3cm
\end{eqnarray}
and it follows from Eqs.(38) and (39) that $\tilde{v}_x=v$ and
$\tilde{v}_y/v=2|\zeta |$. The dependence of $v_y/v$ on
the amplitude of the periodic potential $V(x)=V_0\sin (2\pi x/L)$
is shown in Fig.~2 for different values of periodicity
parameter $L$.

\subsection{Transversal standing wave}

In case of transversal wave the only nonzero component of deformation is
$u_y(x)$. Then in accordance with Eq.(1) we get $V(x)=0$ and $A_x(x)=0$.
Correspondingly, the solutions for the components at ${\bf k}=0$
are
\begin{eqnarray}
\label{41}
&&\varphi _1(x)=\exp \Big( -\int _0^xA_y(x')\, dx'\Big) ,\hskip0.3cm \chi _1 =0,
\\
&&\varphi _2=0,\hskip0.3cm
\chi _2(x)=\exp \Big( \int _0^xA_y(x')\, dx'\Big) .
\end{eqnarray}
The solution for this case was already presented in Sec.~2B, see
Eq.(22).

It should be noted that in reality the longitudinal and transverse
phonon modes in graphene are not completely independent --
there is some mixing between them \cite{manes07}.
The above consideration is fully justified for the phonon
(standing waves) with small $q$.

\section{Screening}

The scalar potential $V(x)$ generated by deformation wave in graphene
is screened by electrons and holes. Screening is the main many-particle
correction to the bare potential which should be taken into account.

For this purpose, using the RPA approximation, we calculate
the loop diagram presenting the polarization operator
\cite{com2}
\begin{eqnarray}
\label{43}
\Pi _0({\bf q})=-i\, {\rm Tr}
\int \frac{d^2{\bf k}\, d\varepsilon }{(2\pi )^3}\;
G({\bf k}+{\bf q},\varepsilon )\, G({\bf k},\varepsilon ),
\end{eqnarray}
where the Green function
\begin{eqnarray}
\label{44}
G({\bf k},\varepsilon )=
\frac{\varepsilon +\mu +v\bsig \cdot {\bf k}}
{(\varepsilon +\mu +i\delta \, {\rm sgn}\, \varepsilon )^2-\varepsilon _k^2}\,
\end{eqnarray}
corresponds to the Hamiltonian of graphene without any
perturbations, $\mu $ is the chemical potential, and $\varepsilon _k=vk$.
For definiteness we assume $\mu >0$. This quantity has been calculated in many papers (see, e.g.,
Refs. [\onlinecite{hwang07,gorbar02,ando06,wunsch06}]) but we present here some intermediate
expressions to discuss a generalization to the anisotropic case.

Substituting Eq.(44) into Eq.(43) and integrating over $\varepsilon $ we find
\begin{eqnarray}
\label{45}
\Pi _0({\bf q})
=\int \frac{d^2{\bf k}}{(2\pi )^2}
\Big\{ [f(\varepsilon _{\bf k+q})-f(\varepsilon _{\bf k})]\,
\frac{\varepsilon _{\bf k+q}^2+\varepsilon _{\bf k}^2 +v^2{\bf k}\cdot {\bf q}}
{\varepsilon _{\bf k+q}(\varepsilon _{\bf k+q}^2-\varepsilon _{\bf k}^2)}
\nonumber \\
+[1-f(\varepsilon _{\bf k})]\,
\frac{-\varepsilon _{\bf k+q}\varepsilon _{\bf k}+\varepsilon _{\bf k}^2+v^2{\bf k}
\cdot {\bf q}}
{\varepsilon _{\bf k}\varepsilon _{\bf k+q}
(\varepsilon _{\bf k+q}+\varepsilon _{\bf k})}
\Big\} ,\hskip0.3cm
\end{eqnarray}
where $f(\varepsilon )=\big\{ \exp [(\varepsilon -\mu )/T]+1\big\} ^{-1}$ is the Fermi
distribution function.
In the limit of small $q\ll \mu /v$ we get
\begin{eqnarray}
\label{46}
\Pi _0({\bf q})\simeq \int \frac{d^2{\bf k}}{(2\pi )^2}
\frac{f(\varepsilon _{\bf k+q})-f(\varepsilon _{\bf k})}
{\varepsilon _{\bf k+q}-\varepsilon _{\bf k}},
\end{eqnarray}
which gives us $\Pi _0(q\to 0)=-\nu (\mu )$,
where $\nu (\varepsilon )=\varepsilon /2\pi v^2$ is the density of
electron states with energy $\varepsilon $ counted from the Dirac point.
The second integral in Eq.(45) includes vacuum screening because
it is nonzero at $\mu =0$, i.e., in graphene without any carriers.
In the limit of $q\to 0$ this contribution to $\Pi _0({\bf q})$
disappears.

\begin{figure}[ptb]
\vspace*{-0.6cm}
\hspace*{-1cm}
\includegraphics[width=1.1\linewidth]{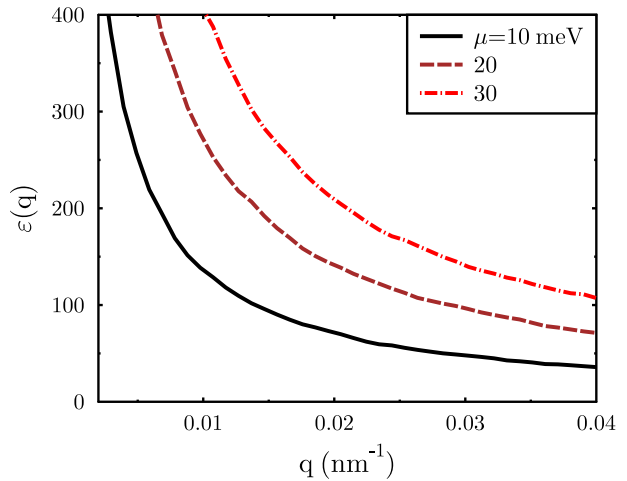}
\caption{Dependence of dielectric constant $\varepsilon $ on $q$ for
different values of chemical potential $\mu $.}
\end{figure}

In frame of the RPA, the screened potential $V(q)$ is related to the bare
potential $V_0(q)$ by $V(q)=V_0+V_0\Pi _0u=V_0/\varepsilon $,
where $u_0(q)$ and $u(q)$ refer respectively to bare and renormalized Coulomb
interaction, and $\varepsilon (q)=1-u_0(q)\, \Pi _0(q)$.
The dielectric constant $\varepsilon (q)$ calculated numerically using Eq.(45)
is presented in Fig.~3. As we see in this figure, the scalar
potential $V(q)$ is substantially suppressed by the screening
since $\varepsilon (q)\gg 1$.

The same dielectric function $\varepsilon (q)$ determines (within the RPA) screening of the
electron-electron interaction, $u(q)=u_0/\varepsilon $.
The value of the effective screening radius is
$R_c=-\big( 2\pi e^2\Pi _0(0)\big) ^{-1}=\big( 2\pi e^2\nu (\mu )\big) ^{-1}$.

In the periodic field, the components of velocity are renormalized,
so that the density of states changes from
$\nu (\varepsilon )=\varepsilon /(2\pi v^2)$ to
$\tilde{\nu }(\varepsilon )=\varepsilon /(2\pi \tilde{v}_x\tilde v_y)$.
Taking into account a variation of the chemical potential in
the periodic field at a constant density of free carriers,
$\tilde{\mu }=\mu \sqrt{\tilde{v}_x\tilde{v}_y}/v$ we find
for the density of states at the Fermi level
$\tilde{\nu }=\mu /(2\pi v\sqrt{\tilde{v}_x\tilde{v}_y})$.
It means that $\tilde{\nu }$ grows with decreasing carrier velocity
as $\tilde{\nu }\sim v/\sqrt{\tilde{v}_x\tilde{v}_y}$
leading to decreasing screening radius $R_c\sim \sqrt{\tilde{v}_x\tilde{v}_y}/v$.
In other words, the periodic potential effectively enhances screening of
the Coulomb interaction between electrons in graphene.

The periodic-field-induced variation of e-e interaction
can affect the many-particle renormalization of the Fermi velocity
\cite{gonzalez99,foster08,elias11} due to modification of the screening.
As follows from presented above estimations, this effect is especially important
when $v/\sqrt{\tilde{v}_x\tilde{v}_y}\gg 1$.

On the other hand, the anisotropy of velocity generated by the
periodic field can essentially modify the renormalization group (RG)
equations of Refs.~[\onlinecite{gonzalez99,elias11}].
Indeed, by calculating the Fock self-energy diagram in the
case of anisotropic spectrum with $\tilde{v}_x\ne \tilde{v}_y$ we find
\begin{eqnarray}
\Sigma ({\bf k})
=\frac{e^2}{4\pi }
\int _{|{\bf k-q}|>k_F}\frac{d^2{\bf q}}{q}\;
\frac{\tilde{v}_x\sigma _x(k_x-q_x)+\tilde{v}_y\sigma _y(k_y-q_y)}
{\tilde{\varepsilon }_{{\bf k-q}}},
\nonumber
\end{eqnarray}
where $k_F$ is the Fermi wave vector, $\tilde{\varepsilon }_k=(\tilde{v}_x^2k_x^2+\tilde{v}_y^2k_y^2)^{1/2}$.
Correspondingly, the velocity correction is
$\delta \tilde{v}_i=\delta \tilde{v}_i^{(1)}+\delta \tilde{v}_i^{(2)}$, where
\begin{eqnarray}
\label{47}
\delta \tilde{v}_i^{(1)}
=\frac{e^2\tilde{v}_i}{4\pi }\int _{|{\bf k-q}|>k_F}
\frac{d^2{\bf q}}{q}\, \frac1{\tilde{\varepsilon }_{{\bf k-q}}},
\end{eqnarray}
\begin{eqnarray}
\label{48}
\delta \tilde{v}_i^{(2)}
=-\frac{e^2\tilde{v}_i}{4\pi k^2}\int _{|{\bf k-q}|>k_F}
\frac{d^2{\bf q}}{q}\, \frac{{\bf k}\cdot {\bf q}}
{\tilde{\varepsilon }_{{\bf k-q}}}.
\end{eqnarray}
In Eq.~(47) we take the limit $k\to 0$
\begin{eqnarray}
\label{49}
\delta \tilde{v}_i^{(1)}
=\frac{e^2\tilde{v}_i}{4\pi }\int _{k_F}\frac{dq}{q}\int _0^{2\pi }
\frac{d\theta }{\sqrt{\tilde{v}_x^2\cos ^2\theta +\tilde{v}_y^2\sin ^2\theta }},
\end{eqnarray}
whereas in (48) we have to take first ${\bf k}=(k,0)$ for
$\delta \tilde{v}_x^{(2)}$ and ${\bf k}=(0,k)$ for $\delta \tilde{v}_y^{(2)}$,
respectively, and after that take the limit of $k\to 0$
\begin{eqnarray}
\label{50}
\delta \tilde{v}_x^{(2)}
=-\frac{e^2\tilde{v}_x}{4\pi }\int _{k_F}
\frac{dq}{q} \int _0^{2\pi }
\frac{\tilde{v}_x^2\cos ^2\theta \, d\theta }
{(\tilde{v}_x^2\cos ^2\theta +\tilde{v}_y^2\sin ^2\theta )^{3/2}},
\nonumber \\
\delta \tilde{v}_y^{(2)}
=-\frac{e^2\tilde{v}_y}{4\pi }\int _{k_F}
\frac{dq}{q} \int _0^{2\pi }
\frac{\tilde{v}_y^2\sin ^2\theta \, d\theta }
{(\tilde{v}_x^2\cos ^2\theta +\tilde{v}_y^2\sin ^2\theta )^{3/2}}.
\end{eqnarray}
The integrals over $q$ in (49),(50) run from $k_F$ to $q_{max}\simeq 1/a_0$.
Since the field-induced renormalization of velocity refers only
to region of small $q<1/L$, we will divide each of these integrals
to the part from $k_F$ to $1/L$ (where, as we found, the spectrum is anisotropic),
and to the part from $1/L$ to $1/a_0$ with $\tilde{v}_x=\tilde{v}_y=v$.

Let us assume for definiteness that for the bare values $\tilde{v}_x/\tilde{v}_y\le 1$.
Using (49),(50) and dividing each of integrals over $q$ in two parts
we find the following many-particle corrections to the velocity
\begin{eqnarray}
\label{51}
&&\delta \tilde{v}_x\simeq \frac{e^2\xi _0}4
+\frac{e^2\tilde{v}_x\xi }{\pi \tilde{v_y}}
\left[ \textsf{K}(m)-\frac{\tilde{v}_x^2}{\tilde{v}_y^2}\, \textsf{R}(m)\right] ,
\nonumber \\
&&\delta \tilde{v}_y\simeq \frac{e^2\xi _0}4
+\frac{e^2\xi }{\pi }
\left[ \textsf{K}(m)-\textsf{P}(m)\right] ,
\end{eqnarray}
where we denoted $\xi _0=\log (L/a_0)$, $\xi =\log (1/k_FL)$,
$\textsf{K}(m)=\int _0^{\pi /2}(1-m\sin ^2\theta )^{-1/2}d\theta $
(complete elliptic integral of the 1st kind\cite{abramowitz}),
$\textsf{P}(m)=\int _0^{\pi /2}(1-m\sin ^2\theta )^{-3/2}
\sin ^2\theta \, d\theta $,
$\textsf{R}(m)=\int _0^{\pi /2}(1-m\sin ^2\theta )^{-3/2}
\cos ^2\theta \, d\theta $, and $m=1-\tilde{v}_x^2/\tilde{v}_y^2$.
As we are interested in the limit of small $k_F$,
the integrals over $q$ are calculated with the logarithmic
precision asssuming $k_F\ll 1/L\ll 1/a_0$.

The first terms in the right hand sides of (51) lead to a constant shift of the
bare values $\tilde{v}_i\to \tilde{v}_i+e^2\xi _0/4$.
Note that in the limit of $\tilde{v}_x=\tilde{v}_y$, Eqs.~(51)
coincide with the ones from Refs. [\onlinecite{gonzalez99,elias11}].

\begin{figure}[ptb]
\vspace*{-0.6cm}
\hspace*{-1cm}
\includegraphics[width=1.1\linewidth]{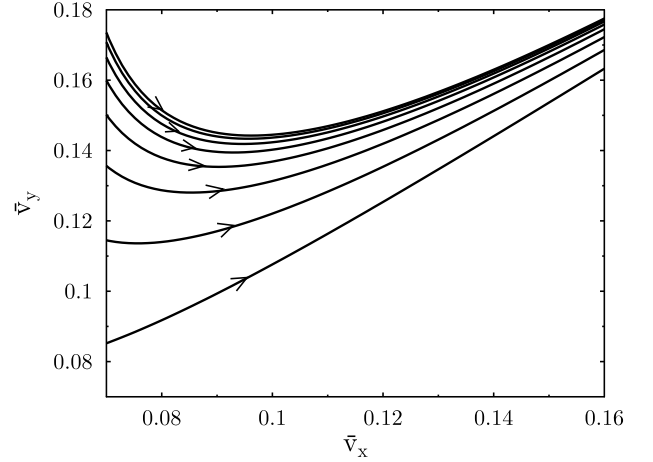}
\caption{The characteristics of RG Eqs.~(52). In the limit
of $\tilde{v}_x\to 0$, all of them go to $\infty $.
The renormalization due to the e-e interaction shifts the initial values of
$\tilde{v}_x+e^2\xi _0/4$ and $\tilde{v}_y+e^2\xi _0/4$ along a certain
characteristic to the right.}
\end{figure}

Using (51) we obtain the following RG equations
\begin{eqnarray}
\label{52}
&&\frac{\partial \tilde{v}_x}{\partial \xi }
=\frac{e^2\tilde{v}_x\xi }{\pi \tilde{v_y}}
\left[ \textsf{K}(m)-\frac{\tilde{v}_x^2}{\tilde{v}_y^2}\, \textsf{R}(m)\right] ,
\nonumber \\
&&\frac{\partial \tilde{v}_y}{\partial \xi }
=\frac{e^2\xi }{\pi }
\left[ \textsf{K}(m)-\textsf{P}(m)\right] .
\end{eqnarray}
The corresponding characteristics in the sector $0<\tilde{v}_x<\tilde{v}_y$ of
($\tilde{v}_x,\tilde{v}_y$) plane are presented in Fig.~4.
As we see, the e-e-interaction-induced renormalization leads to the
effective isotropization of the energy spectrum, which has been broken by
the periodic field.

\section{Plasmons}

First we calculate the real part of polarization operator $\Pi _0({\bf q},\omega )$
for $\omega \ne 0$ assuming that $\varepsilon _k=vk$.
It allows to consider the plasmons in graphene (corresponding to the poles of
dielectric function $\varepsilon (\omega ,{\bf q})$) without external perturbations,
\begin{eqnarray}
\label{53}
{\rm Re}\, \Pi _0({\bf q},\omega )
=-\int \frac{d^2{\bf k}}{(2\pi )^2}
\Big\{ [1-f(\varepsilon _{\bf k+q})]
\nonumber \\ \times
\frac{\varepsilon _{\bf k+q}(\varepsilon _{\bf k+q}-\omega )
+\varepsilon _{\bf k}^2 +v^2{\bf k}\cdot {\bf q}}
{\varepsilon _{\bf k+q}
[(\varepsilon _{\bf k+q}-\omega )^2-\varepsilon _{\bf k}^2]}
\nonumber \\
+[1-f(\varepsilon _{\bf k})]
\frac{2\varepsilon _k^2+\varepsilon _{\bf k}\omega +v^2{\bf k}\cdot {\bf q}}
{\varepsilon _{\bf k}
[(\varepsilon _{\bf k}+\omega )^2-\varepsilon _{\bf k+q}^2]}
\Big\} .
\end{eqnarray}

\begin{figure}[ptb]
\vspace*{-0.6cm}
\hspace*{-1cm}
\includegraphics[width=1.1 \linewidth]{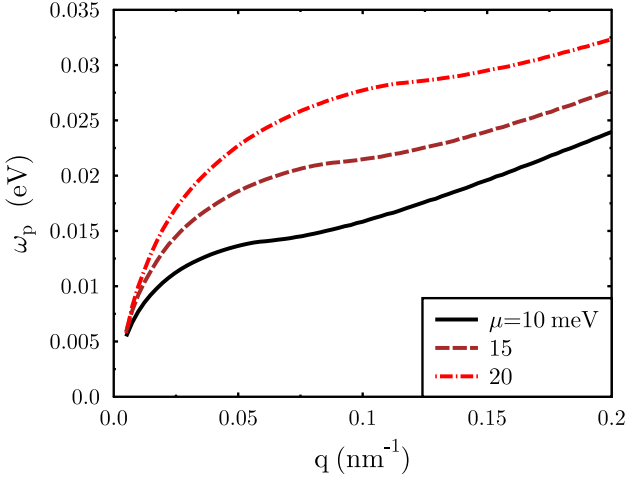}
\caption{Plasmon spectrum $\omega _p(q)$ of graphene for
different values of the chemical potential $\mu $.}
\end{figure}

The dielectric function
$\varepsilon ({\bf q},\omega )=1-\frac{2\pi e^2}{q}\;
{\rm Re}\, \Pi _0({\bf q},\omega )$
can be found using polarization operator (53).
The plasmon spectrum is calculated by solving numerically
equation $\varepsilon ({\bf q},\omega _p)=0$.
It is presented in Fig.~5.
At small $q\ll \mu /v$
the spectrum is $\omega _p(q)\sim \sqrt{q}$ in agreement with
Refs.~[\onlinecite{sarma09,hwang10}].
As we see from Fig.~5, at larger $q$ the plasmon dispersion
is linear with $q$. When $\mu \to 0$, the plasmon spectrum is linear.

As we demonstrated in Sec.~II, the energy spectrum of low-energy excitations
in graphene under the periodic perturbation can be described
by an effective Hamiltonian (35).
After some unitary transformation $T$ this Hamiltonian can be reduced to
the form similar of nonperturbed grapnene
\begin{eqnarray}
\label{54}
T^{-1}\tilde{\mathcal{H}}T
=\tilde{v}_x\sigma _xk_x+\tilde{v}_y\sigma _yk_y.
\end{eqnarray}
with renormalized values of electron velocity.
For example, in the case of longitudinal standing wave,
we have $T=e^{-i\pi \sigma _y/4}$, $\tilde{v}_x=v$ and
$\tilde{v}_y=2v|\zeta |$ with $\zeta $ defined
by Eq.~(40).

The polarization operator ${\rm Re}\, \Pi ({\bf q},\omega )$ in the periodic
field can be found using the same formula (53) after
scaling transformation $k_i=(v/\tilde{v}_i)\tilde{k}_i$ and
$q_i=(v/\tilde{v}_i)\tilde{q}_i$.
Then we find
\begin{eqnarray}
\label{55}
{\rm Re}\, \Pi (\tilde{\bf q},\omega )
=\frac{v^2}{\tilde{v}_x\tilde{v}_y}\;
{\rm Re}\, \Pi _0(\tilde{\bf q},\omega ).
\end{eqnarray}
Since the renormalization of electron velocities is different
for longitudinal and transversal waves, the plasmon spectrum
is different in these cases, too.

As shown before, the plasmon spectrum at $q\to 0$ is proportional
to $\sqrt{q}$. It corresponds to
${\rm Re}\, \Pi _0(q,\omega )\sim q^2/\omega ^2$. Then the scaling
transformation presented above does not change the polarization
operator at $q\to 0$.
Thus, the variation of electron velocity does not affect
the plasmon spectrum at small $q$.

\begin{figure}[ptb]
\vspace*{-0.6cm}
\hspace*{-1cm}
\includegraphics[width=1.1 \linewidth]{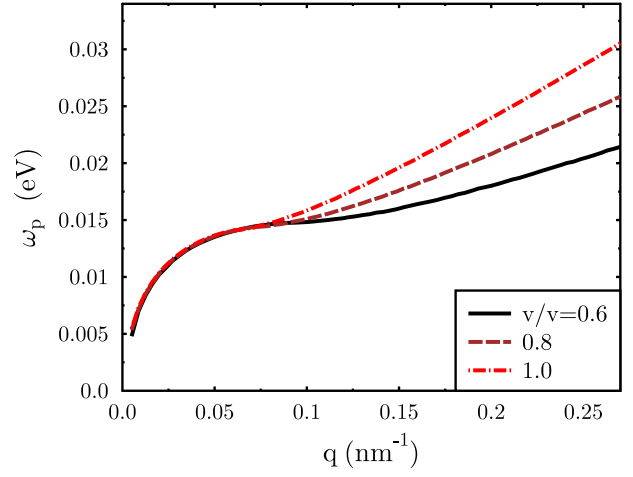}
\caption{Plasmon spectrum $\omega _p(q)$ of graphene under periodic
perturbation for different values of the electron velocity
(without anisotropy like in the case of periodic vector potential).
Here the chemical potential $\mu =10$~meV.}
\end{figure}

In Fig.~6 we present the results of numerical calculation of the plasmon
spectrum for different values of renormalized velocity $\tilde{v}/v$.
It corresponds, e.g., to the presence of periodic field $A_y(x)$.
This figure demonstrates that only the linear part of the spectrum
can be strongly affected by the periodic field.

In the case of longitudinal wave and in the limit of $q\to 0$ we obtain
\begin{eqnarray}
\label{56}
{\rm Re}\, \Pi (\tilde{q},\omega )
\simeq C(\omega )\left( \frac{vq_x^2}{\tilde{v}_y}
+\frac{\tilde{v}q_y^2}{v}\right) .
\end{eqnarray}
It leads to the anisotropy of plasmonic spectrum.

\section{Role of intervalley transitions}

It should be pointed out that our consideration of the plasmonic
spectrum cannot be extended to $q$ of the order of the vector of inverse
lattice. The point is that when one starts from the tight-binding
approximation to describe the electronic structure of graphene,
the Hamiltonian of the Coulomb interaction has the following
form (index $\sigma $ labels sublattices A and B)
\begin{eqnarray}
\label{57}
H_{int}=\sum _{{\bf R}_\sigma {\bf R}'_{\sigma '}}
c^\dag _{{\bf R}_\sigma }c_{{\bf R}_\sigma }\,
u_0({\bf R}_\sigma -{\bf R}'_{\sigma '})\,
c^\dag _{{\bf R}'_{\sigma '}}c_{{\bf R}'_{\sigma '}}
\nonumber \\
=\sum _{{\bf kk'q}\sigma \sigma '}
c^\dag _{{\bf k}\sigma }c_{{\bf k-q},\sigma }u_0({\bf q})\,
c^\dag _{{\bf k}'\sigma '}c_{{\bf k'+q},\sigma '} ,
\end{eqnarray}
where ${\bf k}$ and ${\bf k}'$ are any points in the Brillouin zone.
Considering the states near the Dirac points $\mathcal{K}$
and $\mathcal{K}'$ we can present (57) in a different form
\begin{eqnarray}
\label{58}
H_{int}=
\sum _{{\bf kk'q}\sigma \sigma 'ij}\big[
c^\dag _{{\bf k}\sigma i}c_{{\bf k-q},\sigma i}\, u_0({\bf q})\,
c^\dag _{{\bf k}'\sigma 'j}c_{{\bf k'+q},\sigma 'j}
\nonumber \\
+c^\dag _{{\bf k}\sigma i}c_{{\bf k-q},\sigma j}\, u_0({\bf q-Q})\,
c^\dag _{{\bf k}'\sigma 'j}c_{{\bf k'+q},\sigma 'i}\big] ,
\end{eqnarray}
where $i,j$ labels the valleys, ${\bf k}$ and ${\bf k}'$ are measured
from the corresponding Dirac points, and ${\bf Q}$ is the vector
between the points $\mathcal{K}$ and $\mathcal{K'}$.
This expression shows that for plasmon excitations
with momentum of the order of $Q$, intervalley transitions
should be taken into account, as was pointed out in Ref. [\onlinecite{tudorovskiy10}].

\section{Conclusion}

We considered the variation of electron energy spectrum
near the Dirac point in graphene under periodic perturbation
related to the scalar and vector gauge fields, which
can be generated by the periodic deformations. The
possible source of such deformation fields is a
periodic strain wave like in the case of the ultrasonic wave in
solid. We found that in the general case with both $V\ne 0$ and
${\bf A}\ne 0$ there exist solution for the renormalized
electron velocity, corresponding to the anisotropy of the spectrum.
The problem substantially simplifies in some particular cases.
Namely, for pure longitudinal and pure transverse periodic excitations
the solutions have simple form.

We also considered the screening of the scalar potential
and found that it is strongly suppressed, especially at small $q$.
It means that the main perturbation affecting the electron
velocity is the vector potential ${\bf A}$.

We calculated the plasmon spectrum of collective excitations
in graphene in the presence of periodic excitations.
We found that the plasmon spectrum can be strongly affected
by the periodic field. For the longitudinal excitations,
one appears the anisotropy of plasmon spectrum.

\section*{Acknowledgements}

This work is supported by the Deutsche Forschungsgemeinschaft in Germany,
by the National Science Center as a research project in years 2011 -- 2014
in Poland, and by the ''Stichting voor Fundamenteel Onderzoek der Materie (FOM)'',
which is financially supported  by the ``Nederlandse Organisatie voor Wetenschappelijk
Onderzoek (NWO)''.


\begin{thebibliography}{99}

\bibitem{geim07}
A. K. Geim and K. S. Novoselov, Nature Mater. {\bf 6}, 183 (2007).

\bibitem{castro09}
A. H. Castro Neto, F. Guinea, N. M. R. Peres, K. S. Novoselov, and
A. K. Geim, \rmp {\bf 81}, 109 (2009).

\bibitem{katsbook}
M. I. Katsnelson, {\it Graphene: Carbon in Two Dimensions} (Cambridge Univ. Press, Cambridge, 2012).

\bibitem{acp} P. Avouris, Z. Chen, and V. Perebeinos, Nature Nanotech. {\bf 2}, 605 (2007).

\bibitem{geim09}
A. K. Geim, Science {\bf 324}, 1530 (2009).

\bibitem{bonnacorso10}
F. Bonaccorso, Z. Sun, T. Hasan, and A. C. Ferrari, Nat. Photon. {\bf 4}, 611 (2010).

\bibitem{novoselov11} K. S. Novoselov, \rmp {\bf 83}, 837 (2011).

\bibitem{jablan09}
M. Jablan, H. Buljan, and M. Solja$\check{\rm c}$i\'c,
\prb {\bf 80}, 245435 (2009).

\bibitem{dubinov11}
A. A. Dubinov, V. Ya. Aleshkin, V. Mitin, T. Otsuji, and V. Ryzhii,
J. Phys. Cond. Matter. {\bf 23}, 145302 (2011).

\bibitem{pereira09}
V. M. Pereira and A. H. Castro Neto, \prl {\bf 103}, 046801 (2009).

\bibitem{guineaNP10}
F. Guinea, M. I. Katsnelson, and A. K. Geim,
Nat. Phys. {\bf 6}, 30 (2010).

\bibitem{guinea10}
F. Guinea, A. K. Geim, M. I. Katsnelson, and K. S. Novoselov,
\prb {\bf 81}, 035408 (2010).

\bibitem{vozmediano10}
M. A. H. Vozmediano, M. I. Katsnelson, and F. Guinea,
Phys. Rep. {\bf 496}, 109 (2010).

\bibitem{park08}
C. H. Park, L. Yang, Y. W. Son, M. L. Cohen, and S. G. Louie,
\prl {\bf 101}, 126804 (2008).

\bibitem{park08_N}
C. H. Park, L. Yang, Y. W. Son, M. L. Cohen, and S. G. Louie,
Nature Phys. {\bf 4}, 213 (2008).

\bibitem{tan10}
L. Z. Tan, C. H. Park, and S. G. Louie,
\prb {\bf 81}, 195426 (2010).

\bibitem{LeRoy} M. Yankowitz, J. Xue, D. Cormode, J. D. Sanchez-Yamagishi,
K. Watanabe, T. Taniguchi, P. Jarillo-Herrero, P. Jacquod, and
B. J. LeRoy,     Nature Phys. {\bf 8}, 382 (2012).

\bibitem{com1}
Discussing the effect of vector potential
Tan et al. \cite{tan10} used Lorentz transformation to imaginary
electric field with the subsequent analytic continuation to
the real field. In our method we use the standard quantum mechanics.
Even though our numerical results are in agreement, analitical
formulas are slightly different. In partucular, our Eq.~(22) shows
that the change of electron velocity does not depend on the sign of
periodic perturbation.

\bibitem{wu12}
S. Wu, M. Killi, and A. Paramekanti, \prb {\bf 85}, 195404 (2012).

\bibitem{hwang07}
E. H. Hwang and S. Das Sarma, \prb {\bf 75}, 205418 (2007).

\bibitem{sarma09}
S. Das Sarma and E.H. Hwang, \prl {\bf 102}, 206412 (2009).

\bibitem{tudorovskiy10}
T. Tudorovskiy and S. A. Mikhailov, \prb {\bf 82}, 073411 (2010).

\bibitem{nikitin11}
A. Yu. Nikitin, F. Guinea, T. J. Garc\'ia-Vidal, and L. Mart\'in-Moreno,
\prb {\bf 84}, 161407(R) (2011).

\bibitem{abedinpour11}
S. H. Abedinpour, G, Vignale, A. Principe, M. Polini, W. K. Tse,
and A. H. MacDonald, \prb {\bf 84}, 045429 (2011).

\bibitem{yuan11} S. Yuan, R. Roldan, and M. I. Katsnelson,
Phys. Rev. B {\bf 84}, 035439 (2011).

\bibitem{pyatkovskiy11}
P. K. Pyatkovskiy and V. P. Gusynin, \prb {\bf 83}, 075422 (2011).

\bibitem{suzuura02}
H. Suzuura and T. Ando, \prb {\bf 65}, 235412 (2002).

\bibitem{manes07}
J. L. Ma$\tilde{\rm n}$es, \prb {\bf 76}, 045430 (2007).

\bibitem{tsidilkovskii} I. M. Tsidilkovskii, {\it Band Structure of Semiconductors}
(Pergamon, Oxford, 1982).

\bibitem{com2}
We use matrix Green's functions, which
automatically accounts for correct matrix elements
of e-e interaction and includes possible
vacuum screening of the filled valence bands. This
method has been used for similar calculations
in narrow-gap semiconductors \cite{abrikosov74}.

\bibitem{abrikosov74}
A. A. Abrikosov, Zh. Eksp. Teor. Fiz. {\bf 66}, 1443 (1974)
[Sov. Phys. JETP {\bf 39} (1974)].

\bibitem{gorbar02} E. V. Gorbar, V. P. Gusynin, V. A. Miransky, and
I. A. Shovkovy, Phys. Rev. B {\bf 66}, 045108 (2002).

\bibitem{ando06} T. Ando, J. Phys. Soc. Japan {\bf 75}, 074716 (2006).

\bibitem{wunsch06} B. Wunsch, T. Stauber, F. Sols, and F. Guinea,
New J. Phys. {\bf 8}, 318 (2006).

\bibitem{gonzalez99}
J. Gonzalez, F. Guinea, and M. A. H. Vozmediano,
\prb {\bf 59}, 2474(R) (1999).

\bibitem{foster08}
M. S. Foster and I. L. Aleiner, \prb {\bf 77}, 195413 (2008).

\bibitem{elias11}
D. C. Elias, R. V. Gorbachev, A. S. Mayorov, S. V. Morozov,
A. A. Zhukov, P. Blake, L. A. Ponomarenko, I. V. Grigorieva,
K. S. Novoselov, F. Guinea, and A. K. Geim,
Nature Phys. {\bf 7}, 701 (2011).

\bibitem{abramowitz}
M. Abramowitz and I. A. Stegun, {\em Handbook of Mathematical
Function}, Natl. Bur. Stand. Appl. Math. Ser. 55 (Nat. Bur. Stand.,
Washington, DC, 1964).

\bibitem{hwang10}
E. H. Hwang, R. Sensarma, and S. Das Sarma, \prb {\bf 82}, 195406 (2010).

\end{thebibliography}
\end{document}